\documentclass[a4paper,10pt]{article}
\usepackage{amsmath}
\usepackage{epsfig}
\usepackage{amsfonts}
\usepackage{amssymb}
\newcommand{\dd}[2]{\frac{\partial{#1}}{\partial{#2}}}

\newcommand{\f}[2]{\frac{#1}{#2}}
\newcommand{\w}[1]{\widetilde{#1}}
\title{Sound velocity and acoustic nonlinearity parameter for fluids. Thermodynamic premises.}
\author{Sergey Leble, leble@mifgate.pg.gda.pl\\ Anna Perelomova, anpe@mifgate.pg.gda.pl \\
and Magdalena Ku\'smirek-Ochrymiuk,
ochrymiuk@mifgate.pg.gda.pl\\
\\
  Gda\'nsk University of Technology, \\
ul. G. Narutowicza 11/12, 80-952 Gda\'nsk, Poland\\}
\begin{document}

 \maketitle

\begin{abstract}

New theoretical formulae of the sound velocity and the B/A
nonlinearity parameter for some fluids are presented in the paper.
Semi--ideal and van der Waals models of gas are considered and the
parameters are compared to experiment data. The liquid water model
for equation of state, given by Jeffery and Austin analytical
equation, is considered also and modified on the basis of acoustic
data .
\end{abstract}

\begin{tabular}{rl}
&NOTATIONS:\\
&\\
$x$&space coordinate $[m]$;\\
$t$&time $[s]$;\\
$\rho$&density $[kg/m^{3}]$;\\
$p$&pressure $[N/m^{2}]$;\\
$v$&velocity $[m/s]$;\\
$\eta,\zeta$&viscosity parameters;\\
$\chi$&heat conductivity parameter;\\
$x_{\ast}$,$t_{\ast}$,$\rho_{\ast},p_{\ast},v_{\ast}$&dimensionless variables;\\
$T$&absolute temperature $[K]$;\\
$u$&internal energy per unit mass $[J/kg]$;\\
$\rho_{0}, p_{0},v_{0},e_{0},T_{0}$&unperturbed values;\\
$\acute{\rho},\acute{p}, \acute{v},\acute{e}, \acute{T}$&perturbations;\\
$\beta$&characteristic scale of disturbance;\\
$\alpha$&coefficient responsible for amplitude of acoustic wave;\\
$D_{1}..D_{5}$&dimensionless coefficients in evolution equations;\\
$E_{1}..E_{5}$&coefficients in caloric equation of state;\\
$c$&linear sound velocity $[m/s]$;\\
$B/A$, $C/A$&acoustic parameters of nonlinearity;\\
$s$&entropy $[J \cdot kg^{-1} \cdot K^{-1}]$;\\
$c_{v(p)}$&heat capacity under constant pressure (volume)$[J \cdot kg^{-1} \cdot K^{-1}]$;\\
$R$&the universal gas constant $[J \cdot mol^{-1} \cdot K^{-1}]$;\\
$R_{i}$&individual gas constant $[J \cdot K^(-1) \cdot kg^{-1})]$\\
$\mu$& molar mass $[kg/mol]$;\\
$f_{osc}$&number of oscillation degrees of freedom of a gas molecule;\\
$\theta_{i}$&characteristic temperature of oscillation $[K]$;\\
$\gamma$&adiabatic gas constant ($c_{p}/c_{v}$);\\
$\acute{a}, \acute{b}$&van der Waals constants;\\
$\lambda, \alpha_{1}, v_{B}, T_{B},$&\\
$a, b_{0}, b_{1}, b_{2},$&\\
$A_{0},\Psi_{1},\Psi_{2}, \Psi_{3}$&constans in Jeffery-Austin equation for water;\\
\end{tabular}
\newpage

%=========================================================
\section{Introduction}
%=========================================================

    The experimental researches of some physical properties of different fluids,
such the sound velocity and the acoustic nonlinearity parameters,
are well known and advanced today \cite{Be,Co,Co1, MO}. The
theoretical basis on these problems, still have a lot of aspects
to be studied properly.

  One of the most interesting thing is the connecting of statistical
   thermodynamics and acoustic
studies for gas or liquid, in order to disclose the
micro-properties of a nonlinear propagation medium in a direct
link to the macroscopic one. For example, knowing the structure on
the molecular level for a model medium and comparing its acoustic
properties to real one, we could conclude about a molecular
structure of the fluid. In order to obtain the information we do
not have to solve the system of basic equations. Obviously, we
need both thermic and caloric equations of state, taking into
account the thermodynamical relations between them \cite{Lee}.

    There are used two different representations of the equations of state in
our paper: the Taylor series for thermodynamic variables in a
vicinity of a mechanical equilibrium point \cite{P1,P2,P3} and
some analytical formulas. We start from well-known
ideal/semi-ideal and van der Waals gases, comparing and discussing
the results of the sound velocity and $B/A$ evaluations in both
approaches and experiment. Hereby we consider an analytical
(thermic) equation of state for a liquid water accordingly to a
formula given recently by Jeffery and Austin \cite{JA}.
Application of this equation to find sound velocity $c$ and $B/A$
is realized taking into account the equation for free energy. We
presents a general form of the formula for the sound velocity and
the nonlinear parameter $B/A$, (compare with the popular formula
\cite{Be}) with and without using the mentioned Taylor series.
%The next step
%could be finding a new formula for $B/A$ parameter, in place the
% We expect that both of these approaches give almost the same
%values of $c$, but for $B/A$ and higher order nonlinear
%parameters, it could be not so simple.
(The theme of higher order parameters of fluid is presented in
\cite{Bj}, look also in \cite{LV}. )

    The other interesting question is using the sound velocity, and nonlinear parameters,
to test some new equations of state. We mean that an experimental
value of the sound velocity (and B/A, C/A...) can be compare with
a theoretical one, so then we can except or accept, a new model of
medium. More, we have a mechanism to make some corrections, first
of all we mean the adjusting parameters choice whilst covering
more vide field of applications .

    The following section includes formulating the physical problem
    on the mathematical level,
similar like in \cite{P0,P2}, with using projecting technique.
Widely, this theme was raised earlier in \cite{PLK}.  Next
sections contain general approach to the   fluids parameters and
its adaptation to some individual theoretical models.
%=========================================================
\section{Formulating of mathematical problem}
%=========================================================

  The considered physical problem is the fluid medium (gas or liquid) being under acting
  the acoustic wave. A basic system of the hydrodynamic laws of conservation of momentum,
  energy and mass in one-dimensional flow is given by known equations:
\begin{equation}
\begin{split}
    & \dd{v}{t}+v \dd{v}{x}+\f{1}{\rho} \dd{p}{x}-\left(\f{4}{3}\eta+\zeta\right)\dd{^{2}v}{x^{2}}=0,\\
    &\rho \dd{u}{t}+\rho v \dd{u}{x}+p \dd{v}{x}-\left(\f{4}{3}\eta+\zeta\right)(\dd{v}{x})^{2}-\chi\dd{^{2}T}{x^{2}}=0,\\
    & \dd{\rho}{t}+ \dd{(\rho v)}{x}=0,
\end{split}
\end{equation}
and their simplified forms for nonviscous
and non-heat-conducting fluids:
\begin{equation}
\begin{split}
    & \dd{v}{t}+v \dd{v}{x}+\frac{1}{\rho} \dd{p}{x}=0,\\
    &\rho \dd{u}{t}+\rho v \dd{u}{x}+p \dd{v}{x}=0,\\
    & \dd{\rho}{t}+ \dd{(\rho v)}{x}=0.
\end{split}
\end{equation}
In order to complete the physical problem we use the caloric and the thermic equations
of state. The general forms of these thermodynamic equations, obtained by the Taylor series
of two variables ($p,\rho$) are:
\begin{equation}
    \rho _{0}  \acute{u}=E_{1}  \acute{p}+\frac{E_{2} p_{0} }{\rho _{0} }  \acute{\rho} +\frac{E_{3}}{p_{0} }  \acute{p}^{2} +\frac{E_{4} p_{0} }{\rho _{0}^{2} }  \acute{\rho}^{2}+\frac{E_{5} }{\rho _{0}^{} }  \acute{p} \acute{\rho} +\ldots
\end{equation}
\begin{equation}
    \acute{T}=\f{\vartheta_{1}}{\rho_{0}c_{v}}\acute{p}+\f{\vartheta_{2}p_{0}}{\rho_{0}^{2}c_{v}}\acute{\rho}+\ldots.
\end{equation}
Obviously, we assume the quantities $u,T,p,\rho,v$ have to be
treated as $z=z_{o}+\acute{z}$ (index "zero" means an equilibrium
value of $z$ and "prime" means an addition, caused the sound
wave), so the above formulae was written for the additions only.
Finally, using dimensionless variables:
$$v=\alpha c v_{\ast}, \quad \acute{p}=\alpha c^2 \rho_{0} \acute{p}_{\ast} , \quad \acute{\rho}=\alpha  \rho_{0} \acute{\rho}_{\ast}, \quad x=\beta x_{\ast},\quad t=t_{\ast} \beta/c,$$
where c is the linear sound velocity, given by
\begin{equation}
    c=\sqrt{\frac{p_{0}(1-E_{2})}{\rho_{0}E_{1}}},\label{c}
\end{equation}
$\beta$ means the characteristic scale
of disturbance along $x$ and $\alpha$ is the coefficient responsible to the amplitude
of the acoustic wave, we can formulate problem as the one matrix equation:
\begin{equation}
    \frac{\partial}{\partial t}\Psi +L \Psi =\widetilde{\Psi}+\widetilde{\widetilde {\Psi}}+ O(\alpha^{3}), \qquad \Psi=\left(\begin{array}{c}
    v\\
    \acute{p}\\
    \acute{\rho}\
    \end{array}\right)\label{white}.
\end{equation}\\
This is the nonlinear operator equation of time evolution, where:
%--------------------------------------------------------------------------
\begin{equation}
    \begin{split}
    L=\left(\begin{array}{ccc}
    0 &\dfrac{\partial}{\partial x} &0 \\
    \quad&&\\
    \dfrac{\partial}{\partial x} &0&0 \\
    \quad&&\\
    \dfrac{\partial}{\partial x} &0 &0
    \end{array}\right)\qquad\widetilde{\Psi}=&\quad\alpha \left(\begin{array}{c}
    -v\dfrac{\partial v}{\partial x} + \acute{\rho}\dfrac{\partial \acute{p}}{\partial x}\\
    \quad\\
    -v \dfrac{\partial \acute{p}}{\partial x}+\dfrac{\partial v }{\partial x}\left( \acute{p}D_{1} +\acute{\rho}D_{2}\right)\\
    \quad\\
    -v\dfrac{\partial \acute{\rho}}{\partial x} - \acute{\rho}\dfrac{\partial v}{\partial x}
    \end{array}\right)\\.
    \end{split}
\end{equation}
The asterisks for the variables  were omitted for simplicity, and
$D_{1},D_{2}$ denote dimensionless coefficients, which are
algebraic functions of $E_{1}..E_{5}$ (see \cite{PLK}):
$$D_{1}=\frac{1}{E_{1}}\left(-1+2\frac{1-E_{2}}{E_{1}}E_{3}+E_{5}\right),$$
$$D_{2}=\frac{1}{1-E_{2}}\left(1+E_{2}+2E_{4}+\frac{1-E_{2}}{E_{1}}E_{5}\right).$$
The second--order nonlinearity column $\widetilde{\Psi}$ will
contribute to the $B/A$ nonlinearity parameter. The constants $A$,
$B$ relate to coefficients $D_{j}$ i $E_{j}$ in the following way:
$$  A=[(1-E_{2})/E_{1}]p_{0},\qquad B=-(D_{1}+D_{2}+1)[(1-E_{2})/E_{1}]p_{0}, $$
\begin{equation}
\f{B}{A}=-D_{1}-D_{2}-1.\label{BA}
\end{equation}
Now, let us return to the evolution equation (\ref{white}). The
new application of method of acting projectors was presented also
in \cite{P1}. That is the simple way of separating the leftwards,
rightwards and stationary modes of sound, which are responsible
for the wave propagation effect in 'left' and 'right' directions,
and for other effects (such as "streaming", see\cite{MO}). The
separating of mode is done on the evolution equation level. Acting
by one of mentioned projectors, i.e. unitary, orthogonal operators
$P_{1}$, $P_{2}$: or $P_{3}$ $$P_{1}=\frac{1}{2}\quad\left(
\begin{array}{ccc}
1 &1 &0 \\
1 &1 &0 \\
1 &1 &0
\end{array}\right),\;
P_{2}=\frac{1}{2}\quad\left(
\begin{array}{ccc}
1&-1&0 \\
-1 &1 &0 \\
-1&1 &0
\end{array}\right),\;
P_{3}=\left(
\begin{array}{ccc}
0&0 &0 \\
0 &0 &0 \\
 0&-1 &1
\end{array}\nonumber\right)$$
on the evolution equation gives us a new form of wave equation:
\begin{equation}
    \dd{}{t}P_{i}\Psi+P_{i}L\Psi+P_{i}\tilde{\Psi}=0,\qquad i=1,2,3,
\end{equation}
and introduces the sound velocity $c$.
The simplified version of the equation, neglecting heat conductivity and viscosity of medium, has form:
\begin{equation}
    \dd{\rho_{n}}{t}+c_{n}\dd{\rho_{n}}{x}+\f{\varepsilon}{2}\sum^{}_{i,m}{}Y^{n}_{i,m}\rho_{i}\dd{\rho_{m}}{x}+\ldots=0.
\end{equation}
$$i=1,2,3;\quad m=1,2,3;\quad n=1,2,3;\quad c_{1}=1;\quad c_{2}=-1;\quad c_{3}=0;$$
where $T$ denotes some coefficients matrix, and it is built with
algebraic sums of $D_{1}$ and $D_{2}$, for n=1:
\begin{equation}
\left|\begin{array}{cccc}
Y_{i,m}^{1}&m=1&m=2&m=3\\
i=1&-D_{1}-D_{2}+1&D_{1}+D_{2}-1&0\\
i=2&-D_{1}-D_{2}-3&D_{1}+D_{2}-1&0\\
i=3&-D_{2}-1&D_{2}-1&0
\end{array}\right|\nonumber
\end{equation}
We have to remember that the below form of equation is written for
dimensionless variables.\\

\sloppy

%=========================================================
\section{Velocity of sound and B/A parameter in medium}
 %=========================================================

In order to apply our considerations for some different models of fluids, not only ideal gas,
we use the general expression for the sound velocity. Here, the sound velocity
$c$ takes the part of coefficient in the wave equation.
\begin{equation}
    c^{2}=\left( \dd{p}{\rho} \right)_{S,p=p_{0}, \rho=\rho_{0}}
\end{equation}
Index '$S$' means obviously that entropy is constant,  however in
practice it is enough to assume the reversible adiabatic process
\cite{MO}.

The sound velocity dependence on temperature is written well in
the experimental acoustic papers. So, some experimental figures
show linear dependence $c$ on $T$ for majority of liquids.
However, it must be notice, that the sound velocity grows due
temperature to $74^{o}$C and next becomes smaller for liquid water
case. These special forms of curves $c(T)$-dependence are known in
literature as Willard curves. The peculiarities of water result
from long-range order, strong polarity and strong association of
water molecules.

The previous papers providing experimental data of B/A such
\cite{Be,Co,Co1} and later papers, show that the ratio $B/A$
generally increases slowly with temperature, although there are
some exceptions. The contribution to B/A from temperature changes
is smaller than one due pressure changes.
%---------------------------------------------------------------
\subsection{General formulae for nonlinearity parameters}
%----------------------------------------------------------------

  To find a formula for the sound velocity we need two equations of state: $p=p(\rho ,T)$,
$U=U(\rho, T)$ and the first law of thermodynamics: $dU=TdS -
pdV$, where $S$ means entropy, in form:
$$du=Tds+ \frac{p}{\rho^2}d\rho.$$
($s,u$ are variables expressed per unit mass.)
The differentials $dp$ and $du$ we take as:\\
$$dp=\left(\dd{p}{T}\right)_{\rho}dT+\left(\dd{p}{\rho}\right)_{T}d\rho=\beta_{1}dT+\beta_{2}d\rho$$
$$du=\left(\dd{u}{T}\right)_{\rho}dT+\left(\dd{u}{\rho}\right)_{T}d\rho=\beta_{3}dT+\beta_{4}d\rho$$
Comparing suitably the expressions and introducing them to the
first law of thermodynamics equation, one can obtain a formula:
\begin{equation}
    \frac{dp}{d\rho}=\frac{\beta_{1}}{\beta_{3}}\left(\frac{p}{\rho^{2}}-\beta_{4}+\frac{\beta_{3}\beta_{2}}{\beta_{1}} \right)+\frac{\beta_{1}}{\beta_{3}}T\frac{ds}{d\rho}.
\end{equation}
Next, we can do an assumption of adiabatic process of propagating
sound, so finally:
\begin{equation}
    c^{2}=\frac{\beta_{1}}{\beta_{3}}\frac{p}{\rho^{2}}-\frac{\beta_{1}\beta_{4}}{\beta_{3}}+\beta_{2}.\label{cc}
\end{equation}
where
$$\beta_{1}=\left(\dd{p}{T}\right)_{\rho}\qquad\qquad \beta_{2}=\left(\dd{p}{\rho}\right)_{T} $$
$$\beta_{3}=\left(\dd{u}{T}\right)_{\rho}\qquad\qquad \beta_{4}=\left(\dd{u}{\rho}\right)_{T} $$
That is a new approach, without using of Taylor series for
variables.

If we use the Taylor expansion for relation between pressure and
density, and limit ourselves to quadratic terms, we will get the
expression for $B/A$ nonlinear parameter \cite{Be} in form:
\begin{equation}
    \frac{B}{A}=\frac{\rho}{c_{0}^{2}}\left(\dd{^{2}p}{\rho^{2}}\right)_{\rho_{0},S}=\frac{\rho_{0}}{c_{0}^{2}}\left(\dd{c^{2}}{\rho}\right)_{\rho_{0},S}.\label{B/A}
\end{equation}

%=========================================================
\subsection{Ideal and semi-ideal gas model}
 %=========================================================

  For ideal gas we receive the coefficients in forms: $E_{1}=E_{4}=1/(\gamma-1)$,
$E_{2}=E_{5}=-1/(\gamma-1)$, $E_{3}=0$, $D_{1}=-\gamma$, $D_{2}$,
so using the mentioned equation (\ref{c}) we receive \cite{MO}:
$$c=\sqrt{\frac{\gamma p_{0}}{\rho_{0}}}$$ where
$\gamma=C_{p}/C_{V}$.
$$B/A=\gamma -1,\qquad \qquad  C/A=(\gamma -1)(\gamma -2)$$
  The semi-ideal gas model accepts energy of oscillation of molecule,
and omits energy of rotation, because it is significant for very
low temperatures and light gases only. The model concerns
polyatomic gases only, because for monoatomic ones we have the
same formulas as before.
$$(Mc_{v})_{sid}=(Mc_{v})_{id}+(Mc)_{osc}+\Delta(Mc)_{rot}$$
We use the Einstein -- Planck formulae for vibrational specific heat:
$$(Mc)_{osc}=(MR)\sum_{1}^{f_{osc}}{\left(\frac{\theta_{i}}{T}\right)^2}\frac{e^{\theta_{i}/T}}{\left(e^{\theta_{i}/T}-1\right)^{2}}$$
by using which we get the equation for internal energy for semi -- ideal gases [6]:
\begin{equation}
u=u_{id}+\frac{MR}{\mu}\sum_{1}^{f_{osc}}{\frac{\theta_{i}}{e^{\theta_{i}/T}-1}}
\end{equation}
where $Mc$, $MR$, $\mu$, $f$, $\theta_{i}$ -- vibrational specific
heat, universal gas constant, molar mass, number of degrees of
freedom of a molecule and characteristic temperature in correspond
order. According to (\ref{c}) the sound velocity formula looks
finally:
\begin{equation}
    c^{2}=R_{i}T\left(1+\left(\frac{1}{\gamma -1}+ \sum_{i}{\left(\frac{\theta_{i}}{T_{0}}\right)^{2}e^{\theta_{i}/T_{0}}\left(e^{\theta_{i}/T_{0}}-1\right)^{-2}}\right)^{-1}\right)
\end{equation}
A formula for B/A has a more cumbersome form:
\begin{equation}
\begin{split}
    B/A=&-\frac{E_{2}}{\w{C}_{1}}\left\{-1+\f{1+E_{1}}{\w{C}_{1}}\sum_{i}{(\w{C}_{2i}\w{C}_{3i})} +\w{C}_{0}\right\}+\\
        &-\f{E_{2}}{E_{1}}\left\{1+\f{2}{\gamma-1}+\sum_{i}{(\w{C}_{4i}\w{C}_{5i})}+\f{1+E_{1}}{\w{C}_{1}}\w{C}_{0}-E_{1} \right\}
\end{split}
\end{equation}
where some new symbols mean accordingly:
$$\w{C}_{0}=-\f{1}{\gamma-1}+\sum_{i}{\left\{\w{C}_{2i}(-\w{C}_{3i}-1)\right\}} $$
$$\w{C}_{1}=\f{1}{\gamma-1}+\sum_{i}{\w{C}_{2i}} $$
$$\w{C}_{2i}=\left(\f{\vartheta_{i}}{T_{o}} \right)^{2}\f{\exp(\vartheta_{i}/T_{o})}{(\exp(\vartheta_{i}/T_{o})-1)^{2}} $$
$$\w{C}_{3i}=-2-(\vartheta_{i}/T_{o})+2(\vartheta_{i}/T_{o})\f{\exp(\vartheta_{i}/T_{o})}{(\exp(\vartheta_{i}/T_{o})-1)} $$
$$\w{C}_{4i}=(\vartheta_{i}/T_{o})^{3}\f{\exp(\vartheta_{i}/T_{o})}{(\exp(\vartheta_{i}/T_{o})-1)^{2}} $$
$$\w{C}_{5i}=-1+2\f{\exp(\vartheta_{i}/T_{o})}{(\exp(\vartheta_{i}/T_{o})-1)} $$
The semi--ideal gas model has provided the quite correct data for polyatomic gases, like $CO_{2}, CH_{4}$, for any monoatomic (no oscillations) and diatomic gases we have not interesting difference for both models: semi--ideal and ideal one.
 Some results for a few gases are presented in Table 1 and the temperature
dependence of $c$ and $B/A$ for $CO_{2}$ is shown at Fig.1. and
Fig.2. (more in \cite{PLK}).
\newline
\begin{center}
\begin{footnotesize}
\begin {tabular}{|c|c|c|c|}\hline
Gas&Model of ideal gas& Model of semi--ideal gas&Experimental data\\\hline
             &  $c[m/s]$ & $c[m/s]$&  $c[m/s]$  \\
\hline \hline
$He$         &    972.9  &   972.9  &   971      \\ \hline
$CO_{2}$     &   262.2  & 255.0  &    256.7    \\ \hline
$CH_{4}$     &   434.7  &   431.3  &   430      \\ \hline
\end{tabular}
\newline \\
\textbf{TABLE 1.}  All values in the table are obtained for
$T=273.15K$.
\end{footnotesize}
\newline \\
\end{center}
\begin{center}
        \includegraphics[angle=-90,scale=0.5]
        {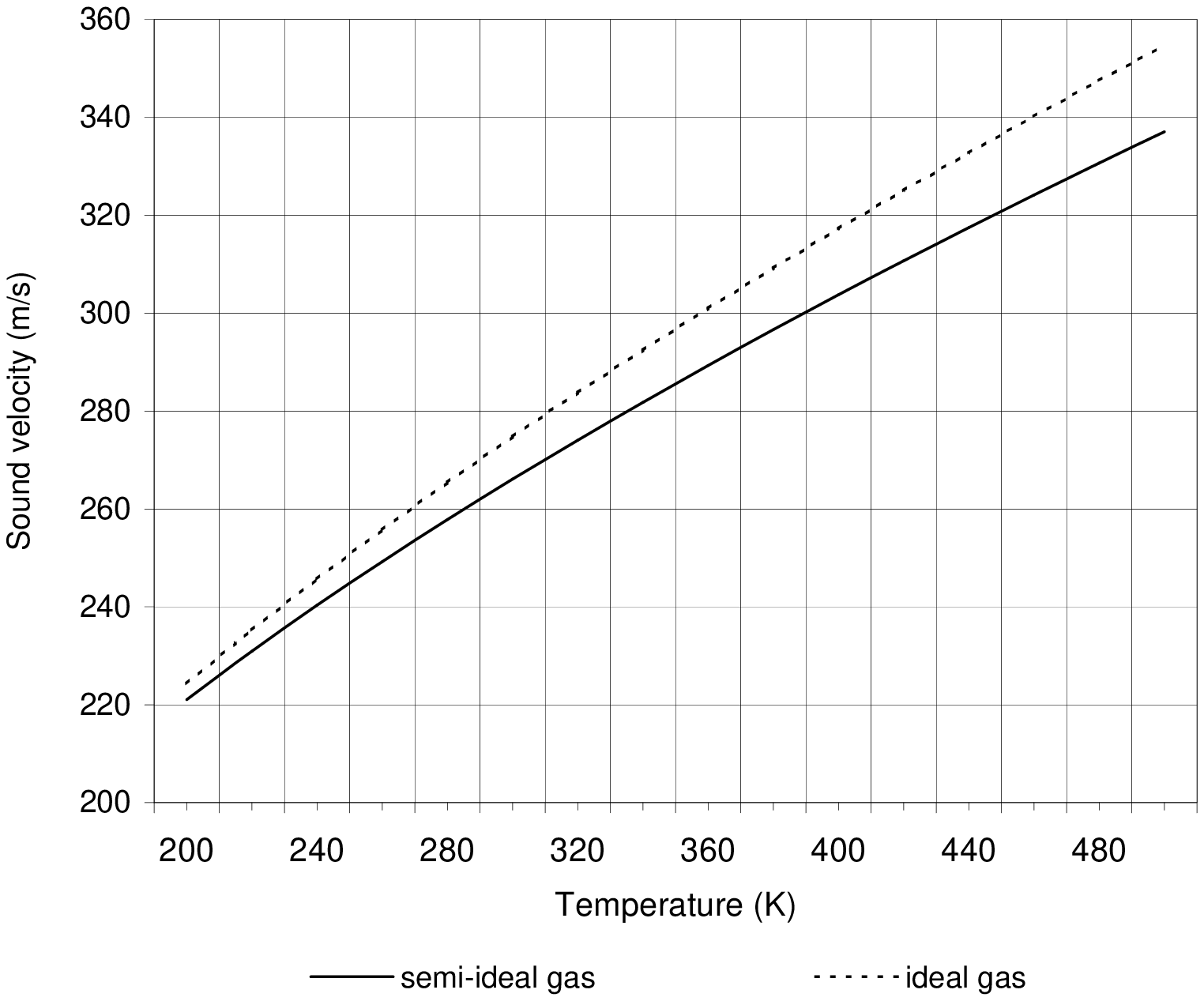}
\label{fig:A1}\\
\textbf{Figure 1.}\\$CO_{2}$.Comparison of temperature dependence of sound velocity \\
for both theoretical models in 200-480 K range of temperature.
\end{center}
\begin{center}
        \includegraphics[angle=0,scale=0.7]{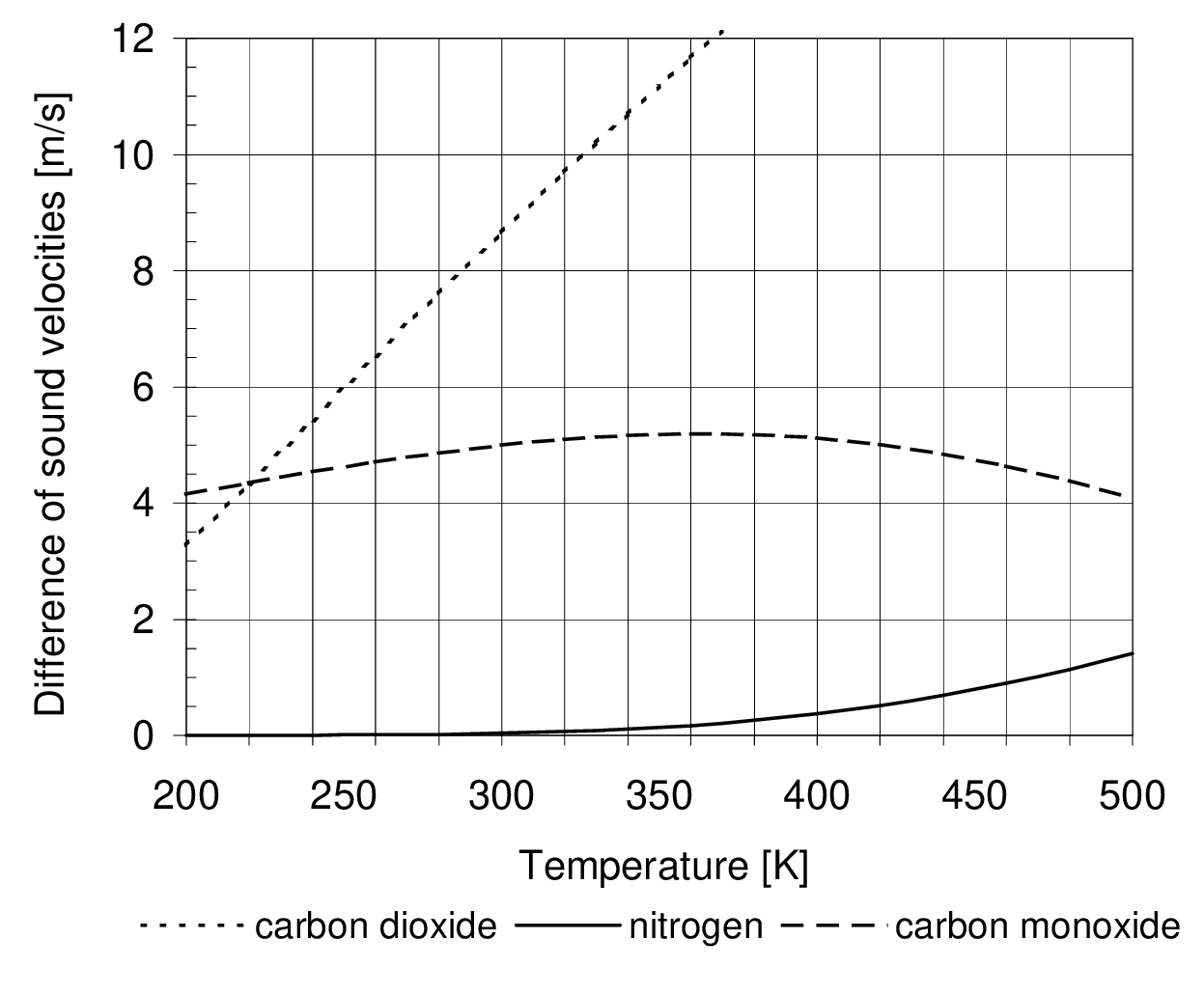}
\label{fig:A1a}\\
\textbf{Figure 1a.}\\
$CO_{2}$ and $CH_{4}$. Differences of sound velocities \\
$c_{id}-c_{sid}$ for $N_{2}$, $CO$ and $CO_{2}$ gases.\\
\end{center}
\begin{center}
        \includegraphics[angle=0,scale=0.7]{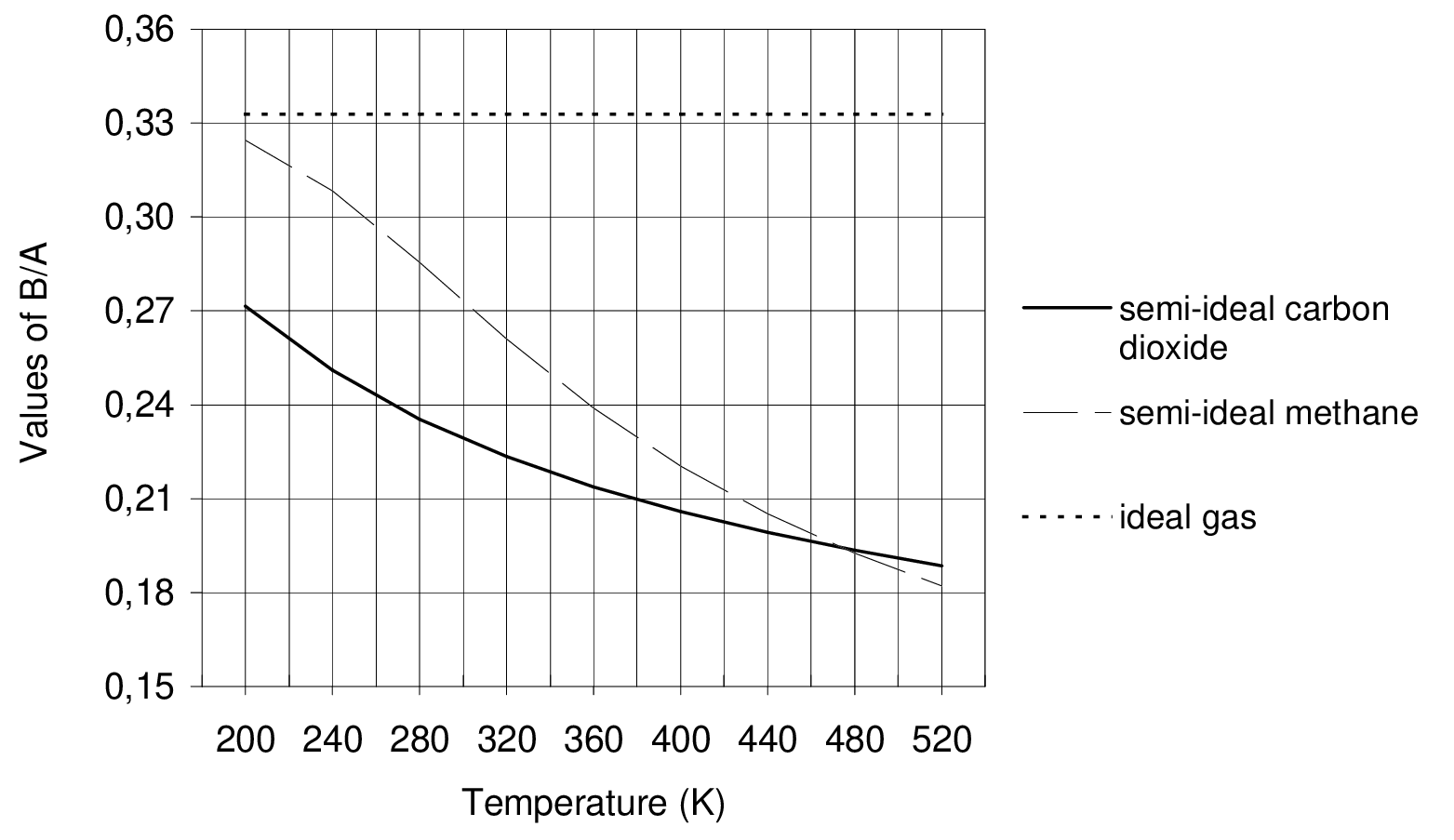}
\label{fig:A2}\\
\textbf{Figure 2.}\\
$CO_{2}$ and $CH_{4}$. Comparison of B/A for both theoretical models \\
in 200-520 K range of temperature. \\
\end{center}
%========================================================
\subsection{Van der Waals gas model}
%=========================================================
Van der Waals gave the famous equation of state for gas model in
1873 year. That is convenient to use the mentioned equation in
form:
\begin{equation}
    p=\frac{\rho R_{i}T}{(1-\acute{b}\rho)}-\acute{a}\rho^{2}\qquad\qquad \text{and}\qquad\qquad e=c_{v}T-\acute{a}\rho,
\end{equation}
where $R_{i}=R/\mu$ means individual constant for gas,
$\acute{a}=a/\mu^{2}, \acute{b}=b/\mu$ - van der Waals constants
and $R_{i},e$ and $c_{v}$ denote some values per unit mass.
Calculating of the sound velocity $c$, according to the  (\ref{c}) gives: $c^{2}=$
\begin{equation}
{{\it R_i}}^{2}T\left (1-\left (2\,{\frac {c_{{v}}a\left
(1-b\rho\right )}{{\it R_i}}}-{\frac {c_{{v}}\left
(p+a{\rho}^{2}\right )b}{\rho\,{\it R_i}}}-{\frac {c_{{v}}\left
(p+a{\rho}^{2}\right )\left (1-b\rho\right )}{{\rho}^{2}{\it
R_i}}}-a\right ){\rho}^{2}{p}^{-1}\right ){c_{{v}}}^{-1}\left
(1-b\rho\right )^{-1}
\end{equation}
and by the new
formula (\ref{cc}), gives a formula:
\begin{equation}
c^{2}=\frac{R_{i}T}{(1-\acute{b}\rho)}\left(\frac{\gamma-1}{(1-\acute{b}\rho)}+1+\frac{\rho\acute{b}}{(1-\acute{b}\rho)}\right)-2\acute{a}\rho;
\end{equation}
Using (\ref{BA}) provides $B/A$ in follow form:
\begin{equation}
\begin{split}
\f{B}{A}=&\left\{R_{i}^{2}p+2R_{i}c_{v}\acute{a}\acute{b}\rho^{3}+R_{i}c_{v}p+\acute{a}R_{i}^{2}\rho^{2}+\right.\\
&+6c_{v}^{2}\acute{a}\acute{b}\rho^{3}+2c_{v}R_{i}\acute{b}p\rho-2c_{v}^{2}\acute{a}\acute{b}^{2}\rho^{4}+2c_{v}^{2}\acute{b}p\rho+\\
&\left.+R_{i}c_{v}\acute{a}\rho^{2}\right\}\left\{(R_{i}p-c_{v}\acute{a}\rho^{2}+2c_{v}\acute{a}\acute{b}\rho^{3}+c_{v}p+R_{i}\acute{a}\rho^{2})c_{v}(-1-b\rho)\right\}^{-1}\\
\end{split}
\end{equation}
  Table 2.
contains a comparison of some results for sound velocity for
$273.15K$. Fig.3. presents the theoretical curve of pressure
dependence for $c$.
\begin{center}
\begin{footnotesize}
\begin {tabular}{|c|c|c|c|c|}\hline
Gas&Model of ideal gas& Laplace formula& Model of van der Waals gas&Experimental data\\\hline
                   & $c[m/s]$& $c[m/s]$& $c[m/s]$ & $c[m/s]$  \\
\hline \hline
$He$               &  970.9  &  971       &   970.7  &  971      \\ \hline
$H_{2}$        & 1259.9  & 1261       &  1259.2  &  1286     \\ \hline
\end{tabular}
\newline \\
\textbf{TABLE 2.}  All values in table are obtained for $T=273.15K$.
\end{footnotesize}
\newline \\
\end{center}
\begin{center}
\includegraphics[angle=0,scale=0.5]{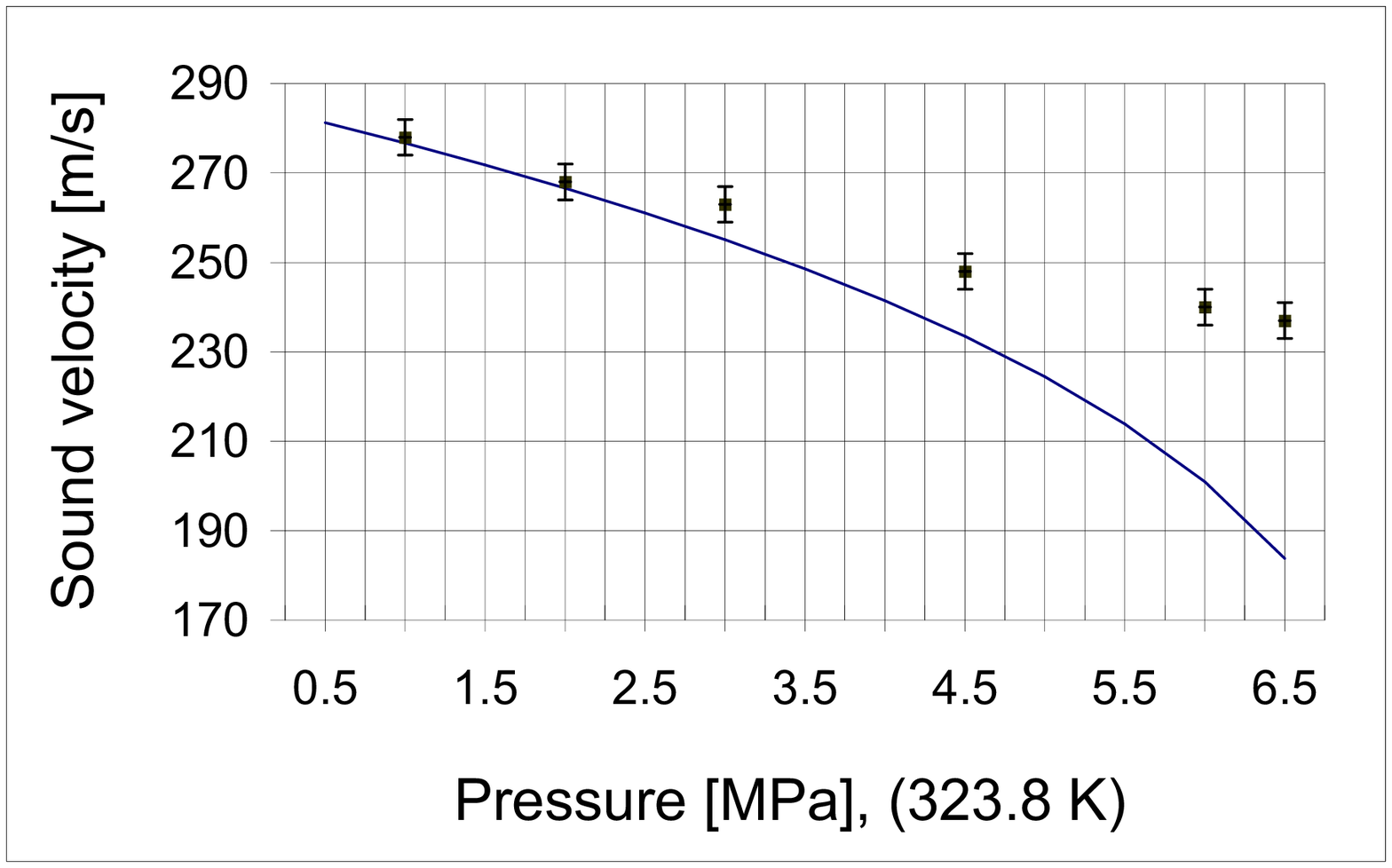}
\label{fig:A3}\\
\textbf{Figure 3.}\\$CO_{2}$. Comparison of theoretical (van der Waals model) \\
and experimental values \cite{BB} of sound velocity for changing
pressure.\\($CO_{2}$, $323.95K$, $0.3MHz$)
\end{center}

  The van der Waals equation of state in its standard form, give
good results for some gases, but is not valid for many of liquids,
in particular for water.
%=========================================================
\subsection{Liquid water model}
%=========================================================
%-----------------------------------------------------------------------------------
\begin{center}\textbf{An analytical equation of state for liquid water}
\end{center}
%-----------------------------------------------------------------------------------
The trials of "build" an analytical equation of state for liquid
water were made. The one of the newest is the Song--Mason--Ihm
equation \cite{SMI} and its modifications, for the polar fluid,
made by Jeffery and Austin \cite{JA}:
\begin{equation}
    \frac{p}{\rho R T}=1-{b_{0}\rho}-\frac{a \rho}{RT}+\frac{\alpha_{1} \rho}{1-\lambda b \rho}
\end{equation}
where $\rho$ is expressed in $(mol/m^{3})$ unit and function $b(T)$ has form:
$$
    b(T)=v_{B}\left(0.25 e ^{1/(2.3 T/T_{B}+0.5)}-b_{1} e ^{2.3 T/T_{B}}+ b_{2}\right)
$$
The constants used by authors of paper have the following values:
$\lambda=0.3159$;
$\alpha_{1}=2.145v_{B}$;
$b_{0}=1.0823v_{B}$;
$b_{1}=0.02774$;
$b_{2}=0.23578$;
$v_{B}=4.1782*10^{-5}m^{3}/mol$;
$T_{B}=1408.4 K$;
$a=0.5542 Pa m^{6}/mol^{2}.$\\
The equation of state for free energy proposed by Jeffery and
Austin for $T>4^{o}C$ has form:
\begin{equation}
F=A_{1}(\rho,T)-RT\Psi(T),
\end{equation}
$$A_{1}=RTlog \rho-RTb_{0}\rho -a\rho-\frac{RT\alpha_{1}}{\lambda b(T)}log(1-\lambda b(T)\rho)-RT(-3log {\Lambda}+1)+A_{0},$$
$$\Psi=\Psi_{1}+\Psi_{2}\frac{T_{B}\lambda b(T)}{T \alpha_{1}}+\Psi_{3}\frac{T_{B}}{T},$$
where: $A_{0}=21.47 kJ/mol,$ $\Psi_{1}=5.13,$ $\Psi_{2}=20.04,$
$\Psi_{3}=2.73,$ a ${\Lambda}$ means temperature wavelength:
$$\Lambda^{2}=\frac{R^{5/3}h^{2}}{2\pi m K_{B}^{8/3}T}.$$\\

In order to find the acoustic wave propagation velocity according
to (\ref{cc}) in discussed medium,
 we make the following calculations for adiabatic process.\\
The known expression for free energy will make possible finding internal energy per unit mass.
In statistical physics: $U=F-T\left(\dd{F}{T} \right)_{V}$, so
$dU=dF-dTF_{T}-TdF_{T}.$
Here, bottom index means partial differential $\dd{F}{T}$.
We can write: $$dF= dA_{1}-RdT\Psi-RT\Psi_{T}dT,\qquad \qquad F_{T}=A_{1T}-R\Psi-RT\Psi_{T},$$
$$dF_{T}=dA_{1T}-Rd\Psi-RdT\Psi_{T}-RT\Psi_{TT}dT,$$ $$dA_{1}=A_{1T}dT+A_{1\rho}d\rho$$

and in the same way like above we obtain the equation:
$$d\rho\left( A_{1\rho}-TA_{1T\rho}+\frac{mp}{\rho^{2}}\right)+dT\left(-TA_{1TT}+2RT\Psi_{T}+RT^{2}\Psi_{TT} \right)=0$$
%-------------------------------------------------------------------------
\begin{center}\textbf{Sound velocity and B/A in liquid water}
\end{center}
%-------------------------------------------------------------------------
and finally:
\begin {equation}
c^{2}=\frac{\left(A_{1\rho}-TA_{1T\rho}+\frac{mp}{\rho^{2}}+\frac{\beta_{2}}{\beta_{1}}TA_{1TT}-\frac{\beta_{2}}{\beta_{1}}2RT\Psi_{T}-\frac{\beta_{2}}{\beta_{1}}RT^{2}\Psi_{TT}\right)\beta_{1}}{TA_{1TT}-2RT\Psi_{T}-RT^{2}\Psi_{TT}}.
\end{equation}
The expression of the nonlinear parameter B/A, received  by using
(\ref{B/A}) has more complicated form, but it is not difficult to
calculate some values using a computer.
%-------------------------------------------------------------------------
\begin{center}\textbf{Results for analytical model of liquid water}
\end{center}
%-------------------------------------------------------------------------

Authors tested Jeffery-Austin analytical equation of state for
liquid water \cite{JA}, being a development of \cite{SMI}. Below
we present some diagrams for the sound velocity and the nonlinear
parameter B/A. The equation seems to be rather sensible for small
changes of constants. Some results of $c$ and $B/A$ for the
Jeffery-Austin equation differ from some experimental data (for
$5-55^{o}C$ temperatures), what was shown
on the following figures (from Fig.4. to Fig.7.). \\
 %<img src="c.doc" alt="0">
%[angle=-90,scale=0.5]

\begin{center}
\includegraphics[angle=0,scale=0.5]{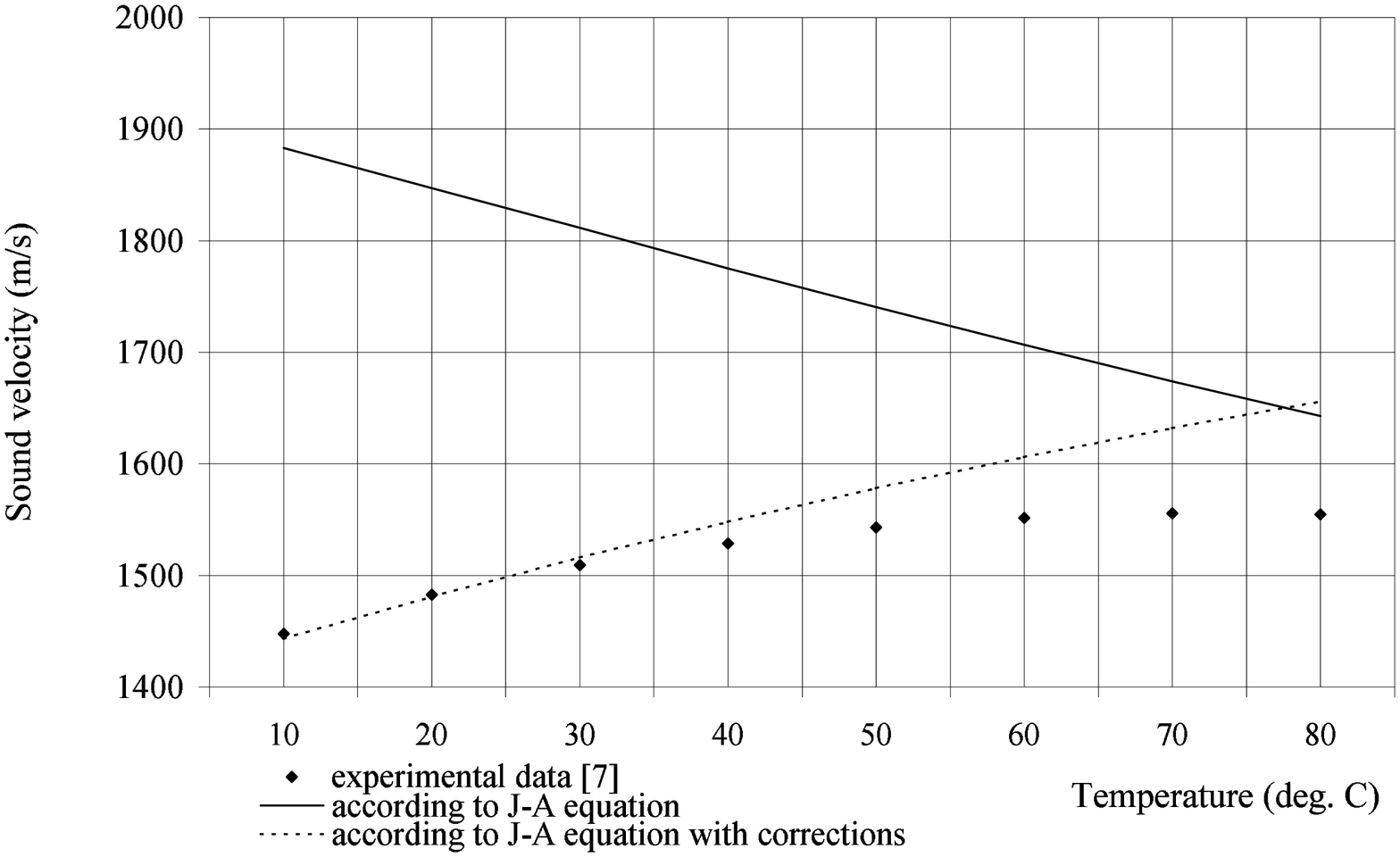}
\label{fig:A4}\\
\textbf{Figure 4.}\\Water. Dependence of sound velocity \\
on temperature in $10^{5}$ Pa pressure.
\end{center}

\begin{center}
\includegraphics[angle=0,scale=0.4]{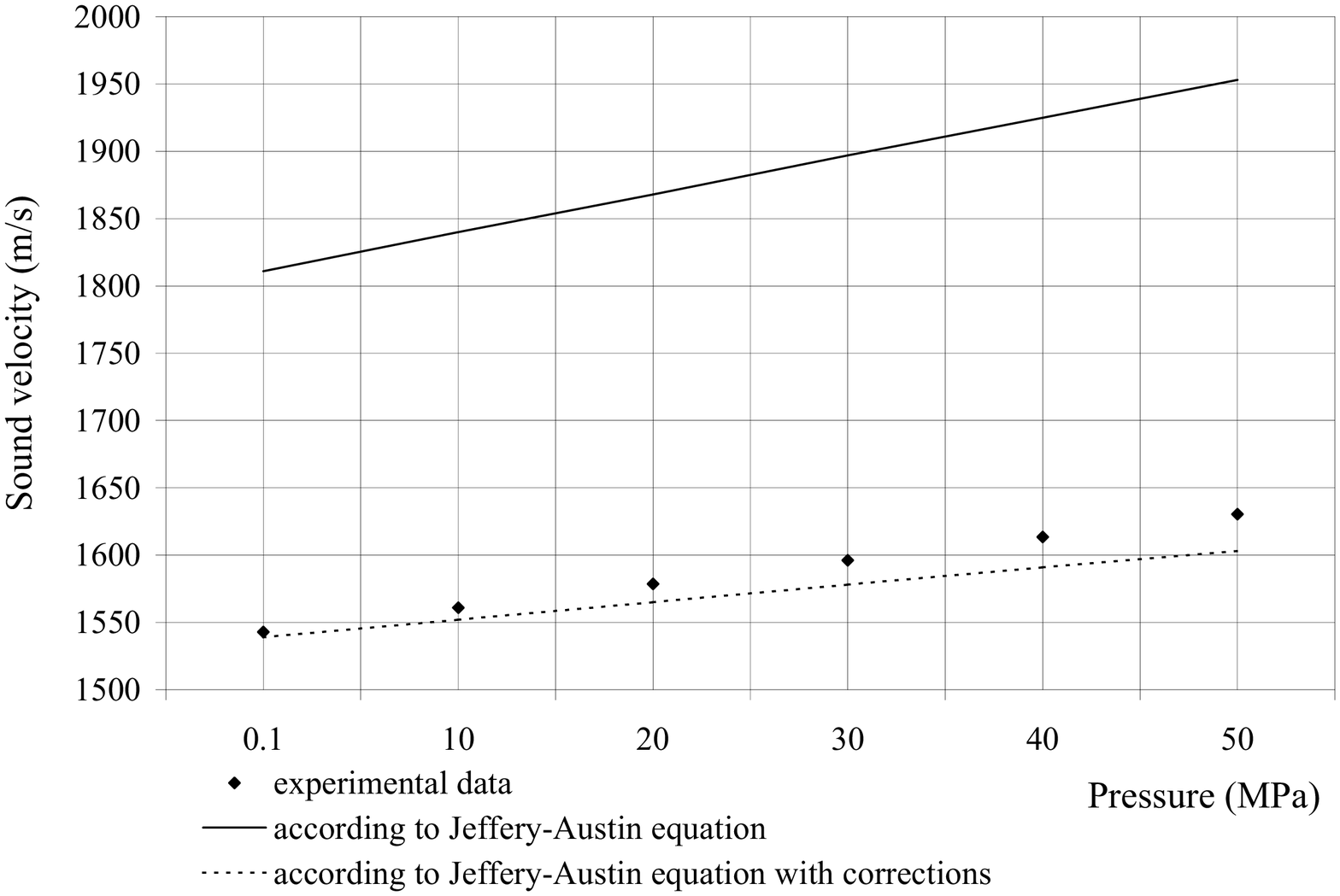}
\label{fig:A5}\\
\textbf{Figure 5.}\\Water. Dependence of sound velocity \\
on pressure, T = 303.15 K.
\end{center}

\begin{center}
\includegraphics[angle=0,scale=0.5]{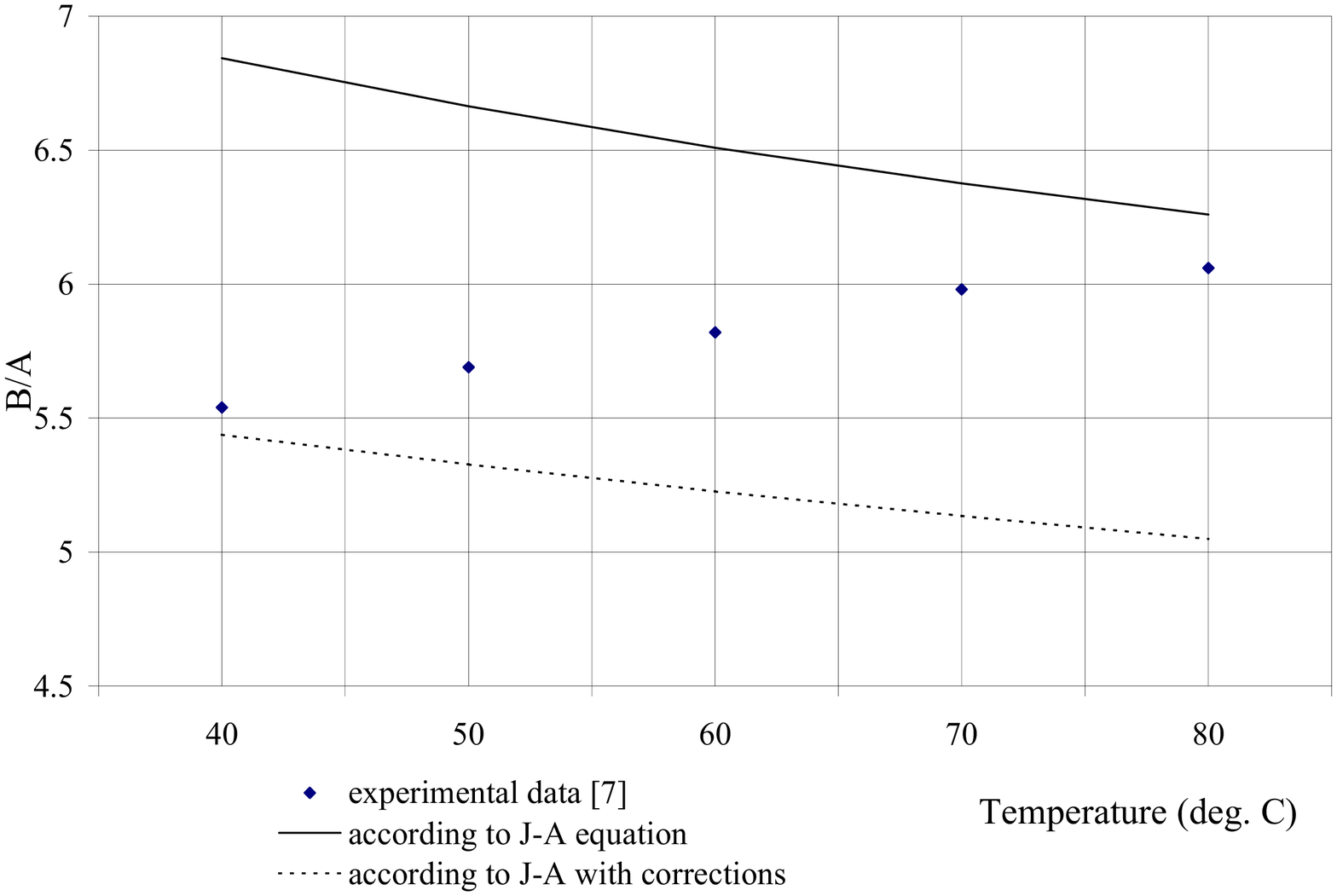}
\label{fig:A6}\\
\textbf{Figure 6.}\\Water. Dependence of $B/A$ parameter \\
on temperature in $10^{5}$ Pa pressure.

\end{center}
\begin{center}
\includegraphics[angle=0,scale=0.5]{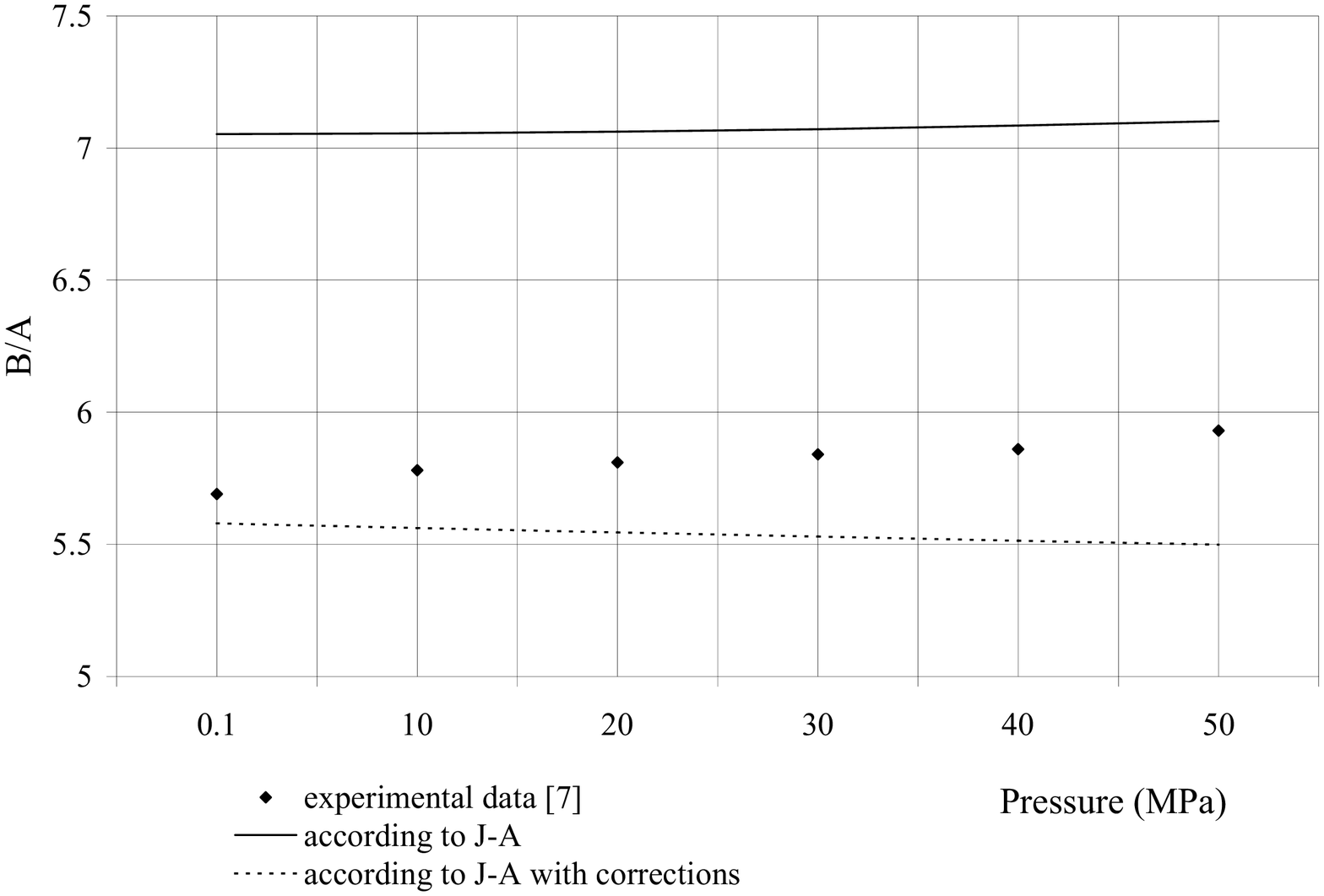}
\label{fig:A7}\\
\textbf{Figure 7.}\\Water. Dependence of $B/A$ parameter \\
on pressure, T = 303.15 K.
\end{center}

  However, the most interesting and worthy to underlining is fact we have a mechanism
   that makes possible testing and even
correcting the equations of state. One of some various possible
corrections is changing the constants: $\lambda=0.244$ instead of
$0.316$, $b_{0}=-0.000026$ instead of $0.000045$ and
$/psi_{2}=22.04$ instead $20.04$. The figures above shows
compatibility that changed model to some experiments. The
presented corrections have an example character only.
%========================================================
\section{Conclusions}
%=========================================================
1. The new presented formula for the sound velocity is better than
the earlier known one, used by the other authors, because of no
necessary using of Taylor series, we have no need to limit
ourselves to some first expressions in that expansion. Although,
the values given by the both of these methods are the same in the
semi-ideal and van der Waals gas model cases, probably we can
expect interesting results of comparison for
 more complicated models of fluids. Undoubtedly, this is the small
 step forward in the theory.

2. Connecting thermodynamic physics and acoustics seems to be an
interesting source of information about considered medium. We
 make a sensitive mechanism to test and correct
theoretical models of various fluids, using of some experimental
data for $c$ and $B/A$. In future, incorporating the links between
statistical physics and thermodynamics,  it could be possible
concluding about intermolecular potentials and molecular structure
of medium from acoustic researches of fluids.

\end{document}